\documentclass[conference]{IEEEtran}
\IEEEoverridecommandlockouts
\usepackage{amsmath,amssymb,amsfonts}
\usepackage{textcomp}

\usepackage[usenames, dvipsnames]{color}
\usepackage{xcolor}
\usepackage{amssymb}
\usepackage{mathtools}
\usepackage{enumerate}
\usepackage{amsmath}
\usepackage{array}
\usepackage{algorithm}
\usepackage{subcaption}
\usepackage{balance}
\usepackage[noend]{algpseudocode}
\usepackage{cite}
\usepackage{flushend}
\usepackage[normalem]{ulem}
\useunder{\uline}{\ul}{}
\usepackage{pdfpages}
\makeatletter
\newcommand*{\rom}[1]{\expandafter\@slowromancap\romannumeral #1@}

\makeatother


\newif\ifComments
\Commentsfalse

\usepackage{graphicx}
\graphicspath{{Images/}}
\DeclareGraphicsExtensions{.pdf,.emf,.png,.eps}

\def\BibTeX{{\rm B\kern-.05em{\sc i\kern-.025em b}\kern-.08em
    T\kern-.1667em\lower.7ex\hbox{E}\kern-.125emX}}
\begin{document}

\title{Data-driven Modelling of Dynamical Systems Using Tree Adjoining Grammar and Genetic Programming\\
\thanks{This research is supported by the Dutch Organization for Scientific Research (NWO, domain TTW, grant: 13852) which is partly funded by the Ministry of Economic Affairs of The Netherlands.}
}

\author{\IEEEauthorblockN{1\textsuperscript{st} Dhruv Khandelwal}
\IEEEauthorblockA{\textit{Department of Electrical Engineering} \\
\textit{Eindhoven University of Technology}\\
Eindhoven, The Netherlands \\
D.Khandelwal@tue.nl}
\and
\IEEEauthorblockN{2\textsuperscript{nd} Maarten Schoukens}
\IEEEauthorblockA{\textit{Department of Electrical Engineering} \\
\textit{Eindhoven University of Technology}\\
Eindhoven, The Netherlands \\
M.Schoukens@tue.nl}
\and
\IEEEauthorblockN{3\textsuperscript{rd} Roland T\'oth}
\IEEEauthorblockA{\textit{Department of Electrical Engineering} \\
\textit{Eindhoven University of Technology}\\
Eindhoven, The Netherlands \\
R.Toth@tue.nl}
}

\maketitle

\begin{abstract}
State-of-the-art methods for data-driven modelling of non-linear dynamical systems typically involve interactions with an expert user. In order to partially automate the process of modelling physical systems from data, many EA-based approaches have been proposed for model-structure selection, with special focus on non-linear systems. Recently, an approach for data-driven modelling of non-linear dynamical systems using Genetic Programming (GP) was proposed. The novelty of the method was the modelling of noise and the use of Tree Adjoining Grammar to shape the search-space explored by GP. In this paper, we report results achieved by the proposed method on three case studies. Each of the case studies considered here is based on real physical systems. The case studies pose a variety of challenges. In particular, these challenges range over varying amounts of prior knowledge of the true system, amount of data available, the complexity of the dynamics of the system, and the nature of non-linearities in the system. Based on the results achieved for the case studies, we critically analyse the performance of the proposed method.
\end{abstract}


\section{Introduction}

Due to rapid increase in computational power over the past decades, there has been a resurgence in the use of Evolutionary Algorithms (EA) across many engineering domains. This trend has also been observed in data-driven modelling (a.k.a identification) of non-linear dynamical systems. While several systematic approaches exist for identification of linear dynamical systems \cite{Ljung1999,van2012subspace}, the problem of non-linear system identification remains challenging. The primary challenge lies in determining a model structure that can capture the dynamics of the true system to the desired level of accuracy, while being parsimonious for human interpretation. Since there exists no systematic approach to choose a suitable model structure for an arbitrary non-linear system, EAs provide a meaningful alternative by allowing exploration of various model structures using evolutionary operations.

Many Symbolic Regression (SR) methods have been proposed in EA literature to estimate the structure of a \textit{static model} from measured data, e.g., see \cite{hoai2002solving,keijzer2003improving}. However, estimating the structure of a \textit{dynamical model} from data poses additional challenges, including:
\begin{itemize}
	\item \textit{Dynamical relations -} at any point in time, the output response of a dynamical system depends not only on the input to the system at that point in time, but also on the \textit{past inputs and outputs} of the system. This complicates the problem of model structure determination.
	\item \textit{Noise - } The presence of noise or unmeasured exogenous signals that influence the dynamics of the system must also be taken into account as a \textit{stochastic} component of the model. The absence of a suitable noise model may lead to biased estimates, as demonstrated in \cite{khandelwal2018on,giordano2016consistency}. 
\end{itemize}

Several EA-based approaches for system identification have been proposed in the literature. A majority of the proposed approaches use EAs to determine the appropriate model complexity within a fixed class of dynamical models (for e.g., see \cite{fonseca1996non, hafiz2018structure, madar2005genetic, kristinsson1992system, rodriguez2004identifying, rodriguez2000use}), or just use EAs to solve the underlying non-linear estimation problem (e.g., see \cite{worden2018evolutionary,aguirre2010prediction}). Alternatively, EAs can be used to build models from a basic set of elements without a prior specification of the structure of the model (e.g., see \cite{schmidt2009distilling, icke2013improving, gray1998nonlinear}). Such approaches are often labelled as \textit{equation discovery} methods. Some closely related work that perform equation discovery without using EA were reported in \cite{brunton2016discovering, kaiser2018discovering} (using sparse regression), in \cite{tanevski2016learning} (using process-based modelling techniques) and in \cite{sahoo2018learning} (using shallow neural networks). 

An approach for data-driven modelling of non-linear systems using Genetic Programming (GP) was presented in \cite{khandelwal2018grammar}. The proposed method makes use of fundamental building blocks to generate model structures, but unlike other equation discovery methods, the construction of new model structures is guided by Tree Adjoining Grammar (TAG) (see \cite{joshi1997tree}). The use of TAG makes the evolutionary search efficient \cite{hoai2002solving} while enabling the modelling framework to function across different model classes with minimal changes in the algorithm. Furthermore, the proposed method also models noise in conjunction with the dynamics of the system. Modelling noise results in smaller variance in the estimated parameters of the model. More crucially, the lack of a suitable noise model may also result in a bias in the estimated model (e.g., see \cite{giordano2016consistency}). The work in \cite{khandelwal2018grammar} was inspired by \cite{hoai2002solving}, where the authors used TAG-guided GP to solve a static symbolic regression problem.

In this paper, we apply the method presented in \cite{khandelwal2018grammar} to three case studies involving real physical systems. The case studies involve modelling of the:
\begin{itemize}
	\item motion dynamics of a pendulum setup,
	\item thermal load-induced deformations of a metal plate,
	\item mechanical behaviour of elastically-coupled electric drives,
\end{itemize}
The objective is to test the performance of the proposed method on modelling tasks that pose prominent challenges encountered in realistic applications. The chosen case studies differ in their characteristics as follows:
\begin{itemize}
	\item while some case studies are well understood in terms of their dynamics (such as the pendulum), other case studies exhibit more complex dynamics (such as unknown non-linearity or delay),
	\item the amount of data available ranges from short (for coupled drives) to large data records (e.g., thermal setup),
	\item the underlying structure of the true system, and the non-linearities involved differ between the case studies.
\end{itemize}
On the basis of the results achieved for the case studies, the performance of the modelling approach is critically analysed. When possible, the results are compared with that achieved by other modelling approaches reported in the literature.

The main contributions of the paper are the following:
\begin{itemize}
	\item we demonstrate that the proposed method can be applied to a diverse set of identification problems, while requiring minimal user interaction or prior knowledge of the dynamical system,
	\item we demonstrate that the results achieved by the proposed method are comparable to those achieved by state-of-the-art non-linear system identification methods that make extensive use of user expertise and/or problem-specific information.
\end{itemize}

\section{Modelling framework}

For completeness, we provide a brief description of the proposed modelling approach. A more detailed account can be found in \cite{khandelwal2018grammar}.

\subsection{Overview}

The proposed method is built upon the following ideas. GP is used to explore model structures that can be generated form a compact set of mathematical operations and terminals. In general, the evolutionary search can be slow and inefficient since GP may also generate model structures that are not well-posed (e.g. non-causal systems) or, in general, not desirable from a modelling perspective. Furthermore, it is not easy to incorporate prior information of the physical system into the evolutionary search systematically. Hence, TAG is used to shape the search-space that can be explored via GP. When no prior information is available, TAG can be used to restrict the search space to an over-arching class of models such as NARMAX (Non-linear Auto-Regressive Moving-Average with eXogenous inputs), which can be used to represent a rich variety of dynamics, both linear and non-linear \cite{billings2013nonlinear}. As special cases, the NARMAX model class includes many commonly used model structures such as ARX and truncated Volterra series \cite{billings2013nonlinear}. When prior information is available, it can be systematically incorporated in the TAG. Within the GP algorithm, a numerical optimization routine is used to estimate model parameters in each iteration of GP. The models are optimized for multiple objective functions (see \ref{sec:MultiObj}) in a pareto-optimal setting.

An overview of the data-driven modelling approach is presented in Algorithm \ref{alg:MOO}. In this Section, we describe the main components of the algorithm.

\begin{algorithm}
	\caption{Multi-objective data-driven modelling using TAG and GP \cite{khandelwal2018grammar}}
	\label{alg:MOO}
	\begin{algorithmic}[1]
		\Require population size $M>0$, number of iterations $L>0$, grammar $G$, probabilities of crossover and mutation $p_c,p_m$
		\State Initialize population $X^{(0)}$, $l=0$, $X^{(-1)}=\{ \}$ \Comment{See \cite{koza1992genetic}}
		\Repeat
			\State Estimate model parameters in $X^{(l)}$ using Least Squares (LS) \Comment{See \cite{khandelwal2018on}} \label{step:parEst}
			\State Compute multi-objective fitness of models in $X^{(l)}$ \label{step:multObj}
			\State Perform non-dominated sorting of populations $X^{(l-1)}$ and $X^{(l)}$ \Comment{See \cite{deb2002fast}}
			\State $X^{(l)} \leftarrow$ first $M$ individuals of the sorted combined population.
			\State Propose new population $X^{(l+1)}$ using crossover and mutation \Comment{See \cite{hoai2003tree}}
			\State $l \leftarrow l+1$
		\Until{$l \le L+1$}
		\Return $X^{(L)}$
	\end{algorithmic}
\end{algorithm}

\subsection{Tree adjoining grammar}

\begin{figure}
		\vspace*{0.2cm}
		\centering
		\begin{subfigure}[t]{0.9\linewidth}
			\includegraphics[scale = 0.41]{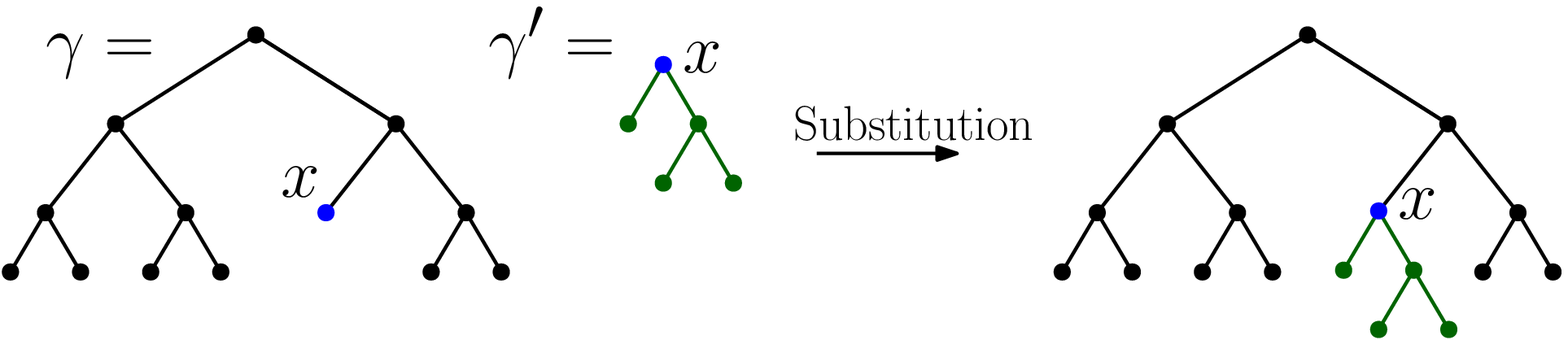}
			\caption{TAG substitution operation.}
			\label{fig:substitution}
		\end{subfigure}
		\\
		\begin{subfigure}[t]{0.9\linewidth}
			\includegraphics[scale = 0.41]{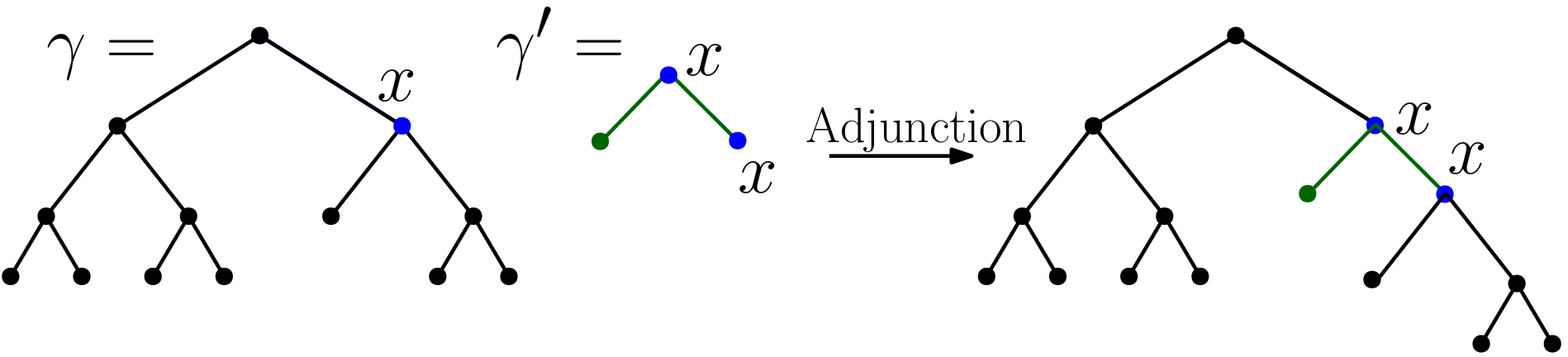}
			\caption{TAG adjunction operation.}
			\label{fig:adjunction}
		\end{subfigure}
		\caption{Illustration of the TAG operations \cite{khandelwal2018grammar}.}
		\label{fig:operations}
\end{figure}

Tree Adjoining Grammar \cite{joshi1997tree} is a tree generating system and defines a \textit{tree language} as the set of all labelled trees that can be generated from a given TAG. The resulting \textit{string language} is defined by the strings appearing at the terminal nodes of all trees in the tree language. A TAG $G$ consists of the following components:
\begin{itemize}
	\item sets of symbols $N$ and $T$ that contain all non-terminal and terminal labels, respectively,
	\item a distinguished start symbol $s \in N$. All trees in the tree language of a grammar $G$ must have $s$ as the label of the root node,
	\item two sets of trees - \textit{initial trees} $I$ and \textit{auxiliary trees} $A$. Initial trees are the most fundamental trees of the tree language. The set of auxiliary trees include trees that can be inserted into an existing tree to obtain a new tree.
\end{itemize}
The \textit{substitution} and \textit{adjunction} operations combine initial and auxiliary trees in specific ways to generate more complex trees that belong to the tree language of a grammar $G$. The substitution operation introduces an initial tree at the terminal nodes of an existing tree. Substitution is valid only if the label of the root node of the initial tree is identical to that of the terminal node of the original tree. The adjunction operation introduces an auxiliary tree to an internal node of an existing tree. Adjunction is valid only if the root node of the auxiliary tree has a label that is identical to that of the chosen internal node. The two operations are illustrated in Fig. \ref{fig:operations}.

\begin{figure}
	\centering
	\includegraphics[width = \linewidth]{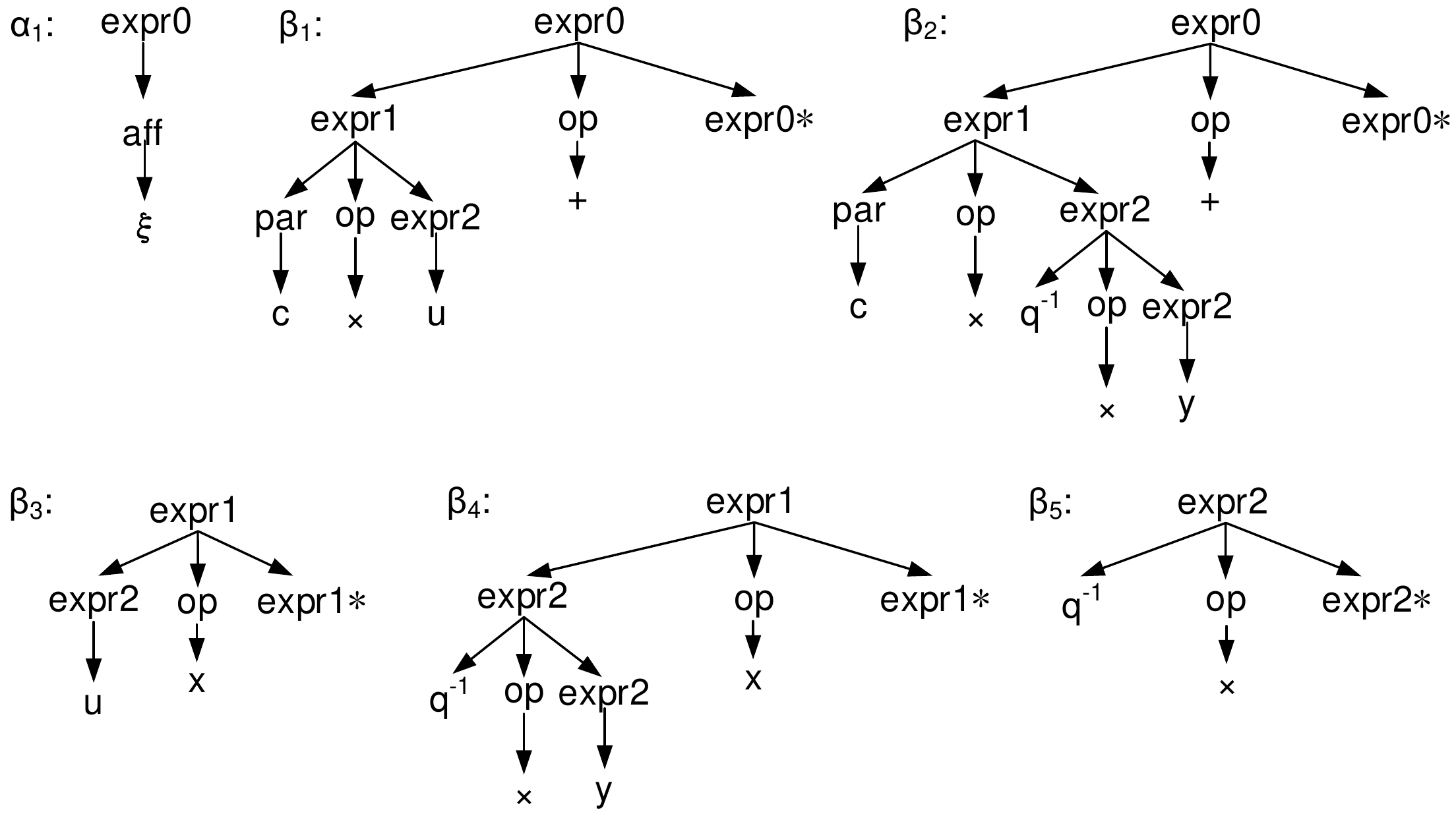}
	\caption{Initial tree $I = \{ \alpha_1 \}$ and auxiliary trees $A = \{ \beta_i \}_{i=1}^5$ of the TAG $G_\mathrm{NARX}$, for details see \cite{khandelwal2018grammar}.}
	\label{fig:TAG_NARX}
\end{figure}

In \cite{khandelwal2018grammar}, the authors proposed a TAG $G_\mathrm{NARX}$ that generates models that belong the discrete-time polynomial NARX model class. The polynomial NARX model class can be represented as
\begin{equation}
	y_k = \sum_{i=1}^p c_i \prod_{j=0}^{n_u} u_{k-j}^{b_{i,j}} \prod_{m=1}^{n_y} y_{k-m}^{a_{i,m}} + \xi_k, \label{eq:NARX}
\end{equation}
where $p$ is the number of model terms, $u_k, y_k \in \mathbb{R}$ are the input and output at time instant $k$, $\xi_k$ is a white noise process, $c_i$ are the model parameters, and $a_{i,m}, b_{i,j} \in \mathbb{Z}_{\ge0}$ are the exponents for the $m^\text{th}$ output factor and the $j^\text{th}$ input factor in the $i^\text{th}$ model term. The initial and auxiliary trees of $G_\mathrm{NARX}$ are shown in Fig. \ref{fig:TAG_NARX}. In \cite{khandelwal2018grammar}, it was also demonstrated that the grammar $G_\mathrm{NARX}$ can be effectively restricted to represent more specific model classes such as linear ARX models. Similarly, it is also possible to extend the grammar to represent a larger class of dynamical systems. Some straightforward extensions to the grammar $G_\mathrm{NARX}$ are used in the case studies presented in Sec. \ref{sec:caseStudies}. 

\subsection{The proposed memetic algorithm}

Genetic Programming is used to explore models defined by the chosen TAG. The exploration takes place across model classes that are included in the TAG, and across varying complexities within a model class. In \cite{hoai2002solving}, the authors proposed a modified GP algorithm based on trees generated by a TAG. The GP algorithm uses modified crossover and mutation operations that ensure that the trees explored during GP iterations belong to the tree language of the chosen TAG. The operations in Step 9 of Alg. \ref{alg:MOO} are based on these modified operations. The sorting and selection operation in Steps 5 and 6 of Alg. \ref{alg:MOO} are based on the non-dominated sorting algorithm proposed in \cite{deb2002fast}. This allows the GP algorithm to evolve non-dominated solutions along the multiple objective functions (see Sec. \ref{sec:MultiObj}).

In each iteration, GP proposes new model structures guided by TAG. The models contain yet-to-be determined parameters (for e.g. $c_i$ in \eqref{eq:NARX}). Since GP is inefficient in optimizing for numerical constants \cite{dabhi2015empirical}, we can use a black-box continuous optimization method such as Covariance Matrix Adaptation - Evolutionary Strategy (CMA-ES)  \cite{hansen2001completely} to estimate model parameters in each iteration. However, as the TAG-generated model classes considered in this paper have either a linear-in-the-parameters or a pseudo-linear structure, we use linear Least Squares (LS) or iterative LS to optimize the model parameters (see Step 3 of Alg. \ref{alg:MOO}).

\subsection{Multi-objective fitness}
\label{sec:MultiObj}

One of the commonly used objective function in system identification is the sum-of-squares of the one-step-ahead error of the model. The one-step-ahead error $e_{k|k-1}$ at time instant $k$ is defined as
\begin{equation}
	e_{k|k-1} \coloneqq \tilde{y}_k - E_\xi(y_{k}|\tilde{y}_{k-1},\tilde{y}_{k-2},\dots, u_{k},u_{k-1},\dots),
\end{equation}
where $\tilde{y}_k$ is the measured output at time-instant $k$ and $E_\xi(y_{k}|y_{k-1},y_{k-2},\dots, u_{k},u_{k-1},\dots)$ is the expectation of the model output at time instant $k$ conditioned on the past measurements $\{y_l\}_{l=1}^{k-1}$ and inputs $\{u_l\}_{l=0}^{k}$. Minimizing the sum-of-squares of the one-step-ahead prediction error yields a computationally efficient estimation of the model parameters, and hence, prediction error is used in Step 3 of Alg. \ref{alg:MOO}. However, prediction error alone is not sufficient to judge the fitness of model since it is in general less sensitive to error arising due to incorrect model structure, especially in the case of non-linear systems \cite{piroddi2003identification}. This is due to the fact that the predicted output is conditioned on the measured data. 

\begin{figure*}
	\centering
	\includegraphics[width = 0.75\linewidth]{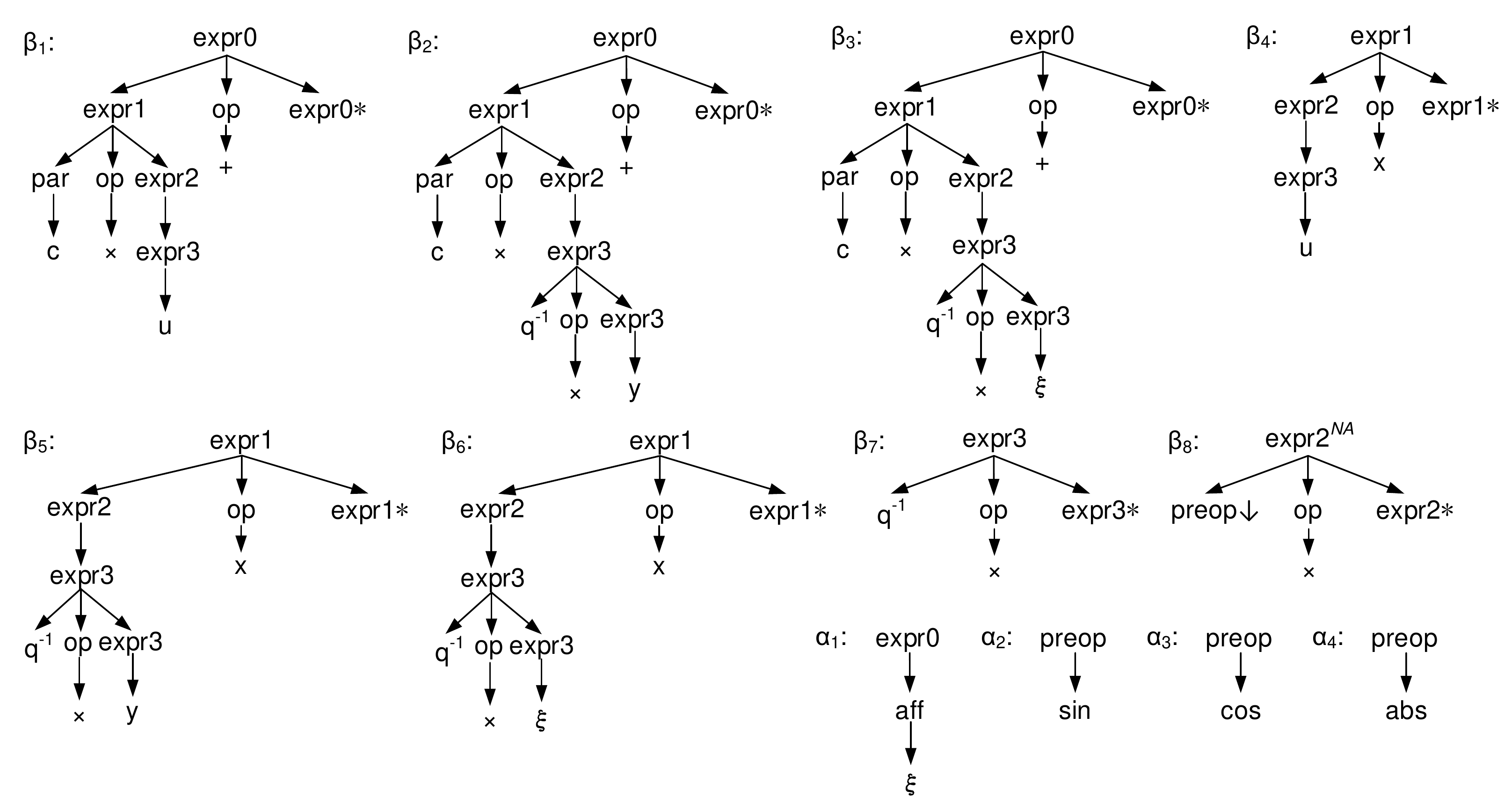}
	\caption{Initial trees $\{\alpha_i\}_{i=1}^{4}$ and auxiliary trees $\{\beta_j\}_{j=1}^{8}$ of the of the over-arching TAG $G$\protect\footnotemark.}
	\label{fig:TAG}
\end{figure*}

The simulation error $\varepsilon_k$ of a model is defined as 
\begin{equation}
	\varepsilon_k \coloneqq \tilde{y}_k - E_\xi(y_{k}),
\end{equation}
where $E_\xi(y_{k})$ is the expected model output w.r.t the distribution of the noise process. Simulation error is typically more sensitive to incorrect model structures since the model output is not conditioned on the measured output. Hence, simulation error and prediction error is used to compute the multi-objective fitness of a model in Step 4 of Alg. \ref{alg:MOO}. For any stochastic candidate model proposed by GP, the corresponding simulation model is computed using the method proposed in \cite{khandelwal2018on}. A third objective function is a complexity measure of a model, defined as the number of coefficients in the model. This allows to evolve models with varying complexities, making it easier for the user to choose the `appropriate' model complexity \textit{a posteriori}.

\section{Case Studies}
\label{sec:caseStudies}

In this section, we present the results achieved by the proposed modelling approach for three case studies. The following quality measures are used to evaluate the results numerically:
\begin{enumerate}
	\item Root Mean Squared prediction error ($\mathrm{RMS}_\mathrm{p}$), computed as
		\begin{equation}
			\mathrm{RMS}_\mathrm{p} = \frac{1}{N-N_t}\sqrt{\sum_{k=N_\mathrm{t}+1}^N e_{k|k-1}^2},
		\end{equation}
		where $N$ is the length of the dataset and $N_t$ is the number of samples discarded to reduce error due to transients.
	\item Root Mean Squared simulation error ($\mathrm{RMS}_\mathrm{s}$), computed as
	\begin{equation}
		\mathrm{\mathrm{RMS}_\mathrm{s}} = \frac{1}{N-N_t}\sqrt{\sum_{k=N_\mathrm{t}+1}^N \varepsilon_k^2}.
	\end{equation}
	\item Best Fit Ratio - prediction ($\mathrm{BFR}_\mathrm{p}$), computed as
	\begin{equation}
		\mathrm{BFR}_\mathrm{p} = 100\left( 1- \frac{\sum_{k=N_\mathrm{t}+1}^N e_{k|k-1}^2}{\sum_{k=N_\mathrm{t}+1}^N (\tilde{y}_k - \mathrm{mean}(\tilde{y}))^2} \right),
	\end{equation}
	where $\mathrm{mean}(\tilde{y})$ is the mean of the measured output.
	\item Best Fit Ratio - simulation ($\mathrm{BFR}_\mathrm{s}$), computed as
	\begin{equation}
		\mathrm{BFR}_\mathrm{s} = 100\left( 1- \frac{\sum_{k=N_\mathrm{t}+1}^N \varepsilon_k^2}{\sum_{k=N_\mathrm{t}+1}^N (\tilde{y}_k - \mathrm{mean}(\tilde{y}))^2} \right).
	\end{equation}
\end{enumerate}
\begin{table}[tb]
\caption{Algorithm hyper-parameters}
\label{tab:hyperparameters}
\centering
\begin{tabular}{l|l}
\hline
\textbf{Hyper-parameter  }              & \textbf{Value }            \\ \hline
GP Population Size            & 100               \\
Maximum GP iterations      & 150               \\
Maximum adjunctions           & 150               \\
Probability of crossover $p_c$ & 1                 \\
Probability of mutation $p_m$  & 0.8               
\end{tabular}
\end{table}

The hyper-parameters of the algorithm are reported in Tab. \ref{tab:hyperparameters}. These hyper-parameters are used for all case-studies presented here. The only hyper-parameter that is tuned to the case study in hand is the grammar used. The choice of grammar is based on (high level) information that is available \textit{a priori}.\footnotetext{The superscript \textit{NA} in auxiliary tree $\beta_8$ refers to a \emph{null adjunction} constraint on the root node, which prohibits the adjunction of any auxiliary tree at that location. See \cite{joshi1997tree} for details.}

The initial and auxiliary trees of the most general TAG used in this paper are shown in Fig. \ref{fig:TAG}. Compared to the TAG in Fig. \ref{fig:TAG_NARX}, the proposed initial and auxiliary trees extend the model class by (i) including polynomial noise factors (i.e. extending to the polynomial NARMAX case), and (ii) including $\sin, \cos$ and $\mathrm{abs}$ functions of monomials. Note that the trigonometric and absolute-value non-linearities are introduced \textit{without} scaling of their arguments, thereby retaining the linear-in-the-parameters structure of the models. The following grammars will be used in the case studies:
\begin{itemize}
	\item The full grammar $G$ with initial trees $\{ \alpha_i \}_{i=1}^4$ and auxiliary trees $\{ \beta_i \}_{i=1}^8$ in Fig. \ref{fig:TAG}.
	\item The grammar $G_\mathrm{trig}$ with initial trees $\{ \alpha_i \}_{i=1}^3$ and auxiliary trees $\{ \beta_i \}_{i=1}^8$ in Fig. \ref{fig:TAG}.
	\item The polynomial NARMAX grammar $G_\mathrm{N}$ with initial trees $\{ \alpha_1 \}$ and auxiliary trees $\{ \beta_i \}_{i=1}^7$ in Fig. \ref{fig:TAG}.
\end{itemize}

\begin{table*}[]
\centering
\begin{tabular}{c|cccccc}
\hline
Case Study                               & Grammar           & $\mathrm{RMS}_\mathrm{p}$ & $\mathrm{RMS}_\mathrm{s}$ & $\mathrm{BFR}_\mathrm{p}$ {[}\%{]} & $\mathrm{BFR}_\mathrm{s}$ {[}\%{]} & \begin{tabular}[c]{@{}c@{}}Estimation\\ time ($\approx$ hrs.)\end{tabular} \\ \hline
\texttt{Pendulum}       & $G_\mathrm{trig}$ & $3.3 \times 10^{-3}$      & $9.4 \times 10^{-2}$      & 99.87                              & 96.15                              & 2                                                                          \\
\texttt{Coupled drives - poly} & $G_\mathrm{N}$    & $4.4 \times 10^{-2}$      & $0.14$                    & 94.59                              & 82.76                              & 2                                                                          \\
\texttt{Coupled drives - general} & $G$               & $4.0 \times 10^{-2}$      & $0.14$                    & 95.06                              & 82.24                             &  2                                                                          \\
\texttt{Thermal setup}  & $G_\mathrm{N}$    & $3.7 \times 10^{-3}$      & $6.2 \times 10^{-3}$      & 99.63                              & 99.28                              & 6.3                                                                         \\
\end{tabular}
\caption{Overview of results.}
\label{tab:overview}
\end{table*}

\subsection{Pendulum}

\begin{figure}
	\centering
	\includegraphics[width = 0.5\linewidth]{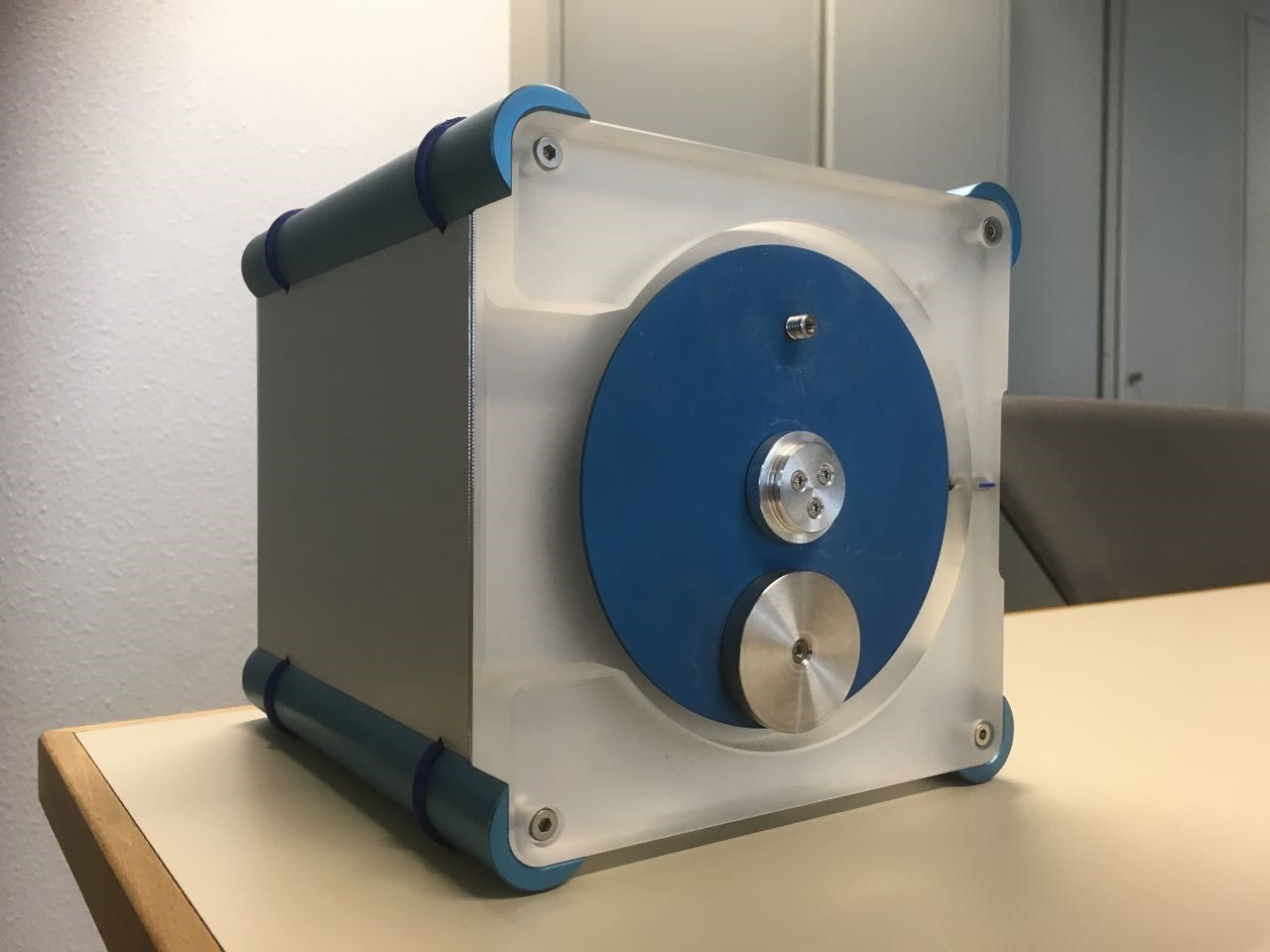}
	\caption{The pendulum setup.}
	\label{fig:pendulum}
\end{figure}

The pendulum system is implemented as a voltage-controlled DC motor attached to a circular disc with an unbalanced mass. The system is shown in Fig. \ref{fig:pendulum}. A torque can be applied to the mass by applying an excitation signal to the motor and the angular displacement is measured in radians. The sampling frequency of the system is 50 Hz. Three datasets are collected and used as follows:
\begin{itemize}
	\item \textit{Estimation dataset} is used to estimate the model parameters in step 3 of Alg. \ref{alg:MOO}.
	\item \textit{Validation dataset} is used to compute the multi-objective fitness of models in step 4 of Alg. \ref{alg:MOO}.
	\item \textit{Test dataset} is used to estimate the generalization error of the final pareto-set of models returned by Alg. \ref{alg:MOO}.
\end{itemize}
The excitation signals were uniformly distributed in the range $[-A_{\mathrm{max}},A_{\mathrm{max}}]$, with the maximum amplitude $A_{\mathrm{max}}$ being 15, 10 and 12 V for the estimation, validation and test datasets respectively. All excitation signals were subsequently low-pass filtered using a discrete-time filter with transfer function
\begin{equation}
	H(z^{-1}) = \frac{0.1z^{-1}}{1 - 0.9z^{-1}},
\end{equation}
and bandwidth $0.85$ Hz. Each input sequence has length $N = 1000$ samples, where the first $N_\mathrm{t} = 50$ samples are set to 0 to remove any transients. The estimation and test dataset, due to their higher amplitudes, include the flipping of the pendulum by more than $\pi$ radians, while the validation dataset does not include the flipping behaviour. This is done to ensure that the system is sufficiently excited to exhibit non-linearities.

\begin{figure}
		\vspace*{-0.2cm}
		\begin{subfigure}[b]{\linewidth}
			\centering
			\includegraphics[scale = 0.7]{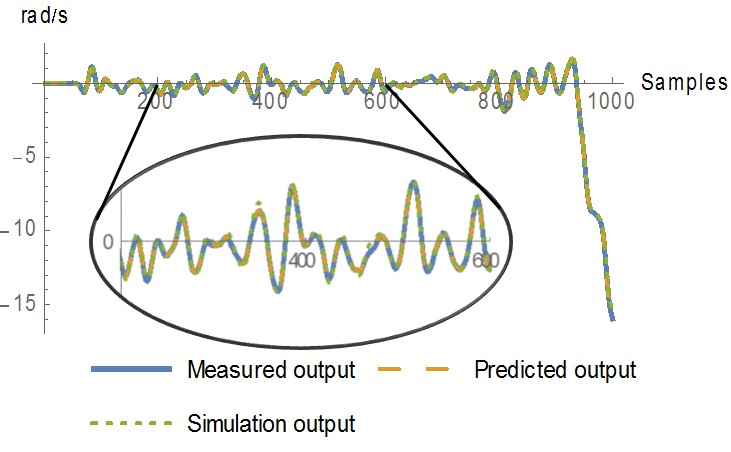}
			\caption{Measured output and the prediction and simulation of the estimated 8 parameter model.}
			\label{fig:pendulumModelResponse}
		\end{subfigure}
		\\
		\begin{subfigure}[b]{0.45\linewidth}
		\vspace*{0.2cm}
			\includegraphics[scale = 0.95]{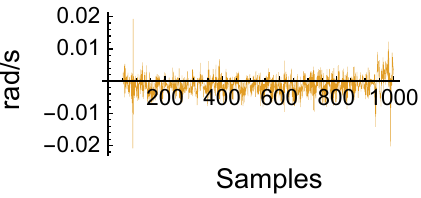}
			\caption{Prediction error.}
			\label{fig:pendulumPredErr}
		\end{subfigure}
		~~~%
		\begin{subfigure}[b]{0.45\linewidth}
			\includegraphics[scale = 0.95]{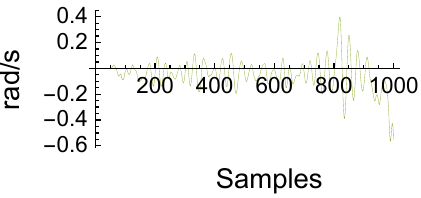}
			\caption{Simulation error.}
			\label{fig:pendulumSimErr}
		\end{subfigure}
		\vspace*{0.2cm}
		\\
		\begin{subfigure}[b]{0.45\linewidth}
			\includegraphics[scale = 0.95]{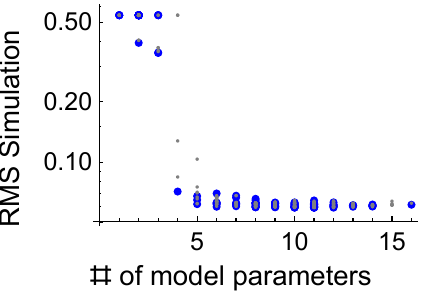}
			\caption{Pareto front - simulation vs complexity.}
			\label{fig:pendulumSim}
		\end{subfigure}
		~~~~%
		\begin{subfigure}[b]{0.45\linewidth}
			\includegraphics[scale = 0.95]{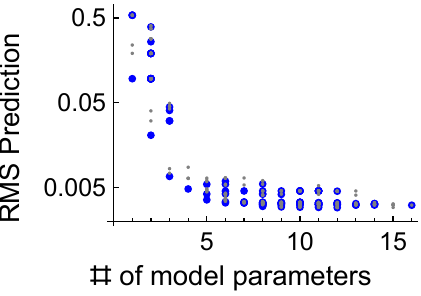}
			\caption{Pareto front - prediction vs complexity.}
			\label{fig:pendulumPred}
		\end{subfigure}
		\caption{Numerical results for the pendulum case-study.}
		\label{fig:pendulumResults}
		\vspace*{-0.4cm}
\end{figure}

Since the system exhibits rotational dynamics, it is reasonable to include trigonometric functions such as sine and cosine in the grammar. Hence, we use $G_\mathrm{trig}$ to define the model set. The results obtained from the identification algorithm is presented in Fig. \ref{fig:pendulumResults}. The best quality measures achieved by models that belong to the pareto-front on the test dataset are reported in Tab \ref{tab:overview}. In Fig. \ref{fig:pendulumModelResponse}, the simulation and prediction error of estimated model with 8 parameters is plotted along with the measured output of the test dataset. It can be observed that the model captures the flipping behaviour (seen at the end of the dataset) in both prediction and simulation. The 2-D projections of the final pareto-fronts are depicted in Fig. \ref{fig:pendulumSim} and \ref{fig:pendulumPred}. The pendulum case study illustrates that the modelling approach could automatically build model structures to describe the flipping behaviour of the system without prior knowledge. This would not be possible within the paradigm of linear time-invariant systems.

\subsection{Coupled drives}

\begin{figure}
	\centering
	\includegraphics[width = 0.7\linewidth]{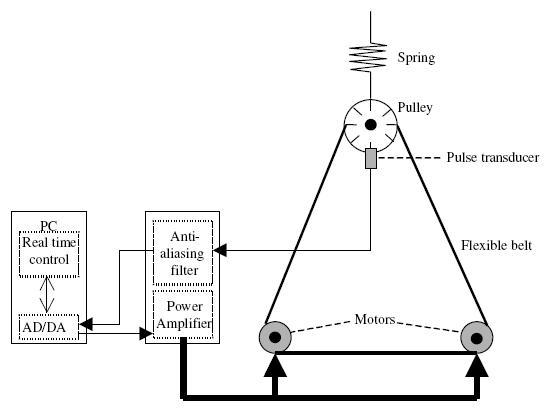}
	\caption{The coupled electric drives setup \cite{wigren2017coupled}.}
	\label{fig:coupledDrive}
\end{figure}

The coupled drives dataset was proposed as a benchmark dataset for non-linear system identification in \cite{wigren2017coupled}. The setup consists of two electric motors driving a flexible belt around a pulley. The setup is illustrated in Fig. \ref{fig:coupledDrive}. The input signal in the dataset is the sum of voltages applied across the two motors, and the output is the pulley velocity as measured by a pulse-counting sensor. The main modelling challenges are (i) short datasets of 500 samples each, and (ii) the sensor is insensitive to the sign of the velocity, resulting in an absolute value non-linearity. The following datasets were used
\begin{itemize}
	\item \textit{Estimation dataset} - input $u11$ and output $z11$ (see \cite{wigren2017coupled}),
	\item \textit{Validation dataset} - input $u12$ and output $z12$ (see \cite{wigren2017coupled}).
\end{itemize}

\begin{figure}
		\begin{subfigure}[b]{\linewidth}
			\centering
			\includegraphics[scale = 0.85]{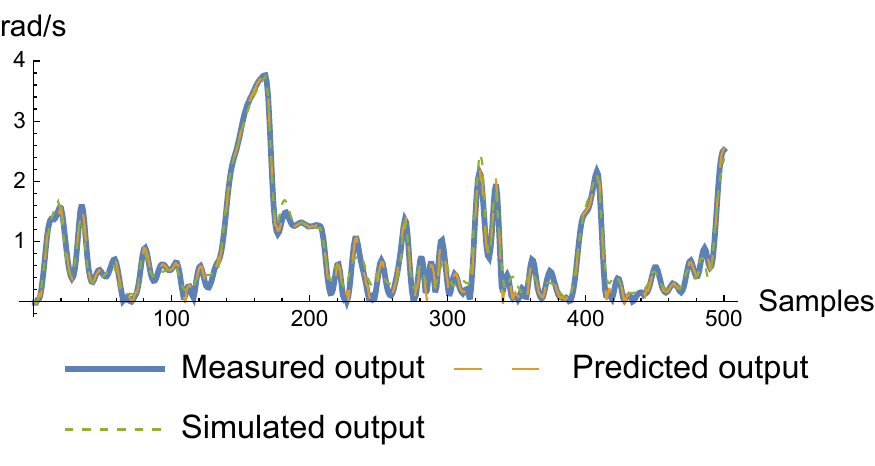}
			\caption{Measured output and the prediction and simulation of the pareto-optimal model with 11 parameters for the second experiment.}
			\label{fig:CDModelResponse}
		\end{subfigure}
		\\
		\begin{subfigure}[b]{0.45\linewidth}
		\vspace*{0.2cm}
			\includegraphics[scale = 0.95]{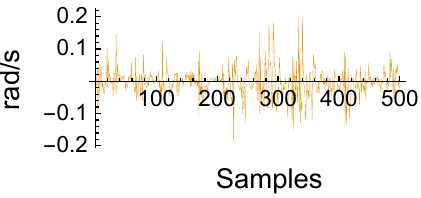}
			\caption{Prediction Error.}
			\label{fig:CDPredErr}
		\end{subfigure}
		~~~%
		\begin{subfigure}[b]{0.45\linewidth}
			\includegraphics[scale = 0.95]{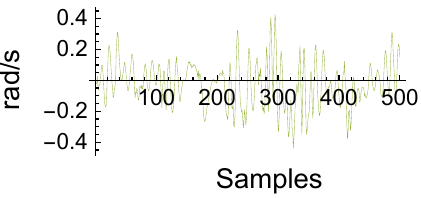}
			\caption{Simulation error.}
			\label{fig:CDSimErr}
		\end{subfigure}
		\vspace*{0.2cm}
		\\
		\begin{subfigure}[b]{0.45\linewidth}
			\includegraphics[scale = 0.95]{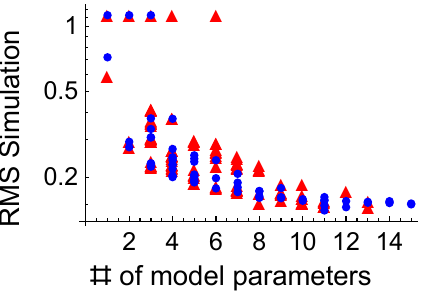}
			\caption{Pareto front for the identification with grammars $G_\mathrm{N}$ (blue) and $G$ (red) - simulation vs complexity.}
			\label{fig:CDSim}
		\end{subfigure}
		~~~%
		\begin{subfigure}[b]{0.45\linewidth}
			\includegraphics[scale = 0.95]{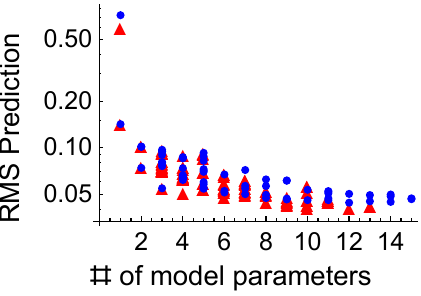}
			\caption{Pareto front for the identification with grammars $G_\mathrm{N}$ (blue) and $G$ (red)  - prediction vs complexity.}
			\label{fig:CDPred}
		\end{subfigure}
		\caption{Numerical results for the coupled drives case-study.}
		\label{fig:CDResults}
		\vspace*{-0.4cm}
\end{figure}

Two identification experiments are performed for this setup. In the first experiment, the polynomial NARMAX grammar $G_\mathrm{N}$ is used. This is referred to as \texttt{Coupled drives - poly} in Tab. \ref{tab:overview}. In the second experiment, the complete grammar $G$ is used as it contains the $\sin, \cos \text{ and } \mathrm{abs}$ functions, which are appropriate for the given case study. This experiment is referred to as \texttt{Coupled drives - general} in Tab. \ref{tab:overview}. The results obtained are depicted in Fig. \ref{fig:CDResults}. The 2-D projections of the pareto-fronts of the two experiments are plotted in Fig. \ref{fig:CDSim} and \ref{fig:CDPred}. In both plots it can be observed that the pareto-fronts of identified models are fairly similar, with the models identified using grammar $G$ performing slightly better in general. We know form Weirstrass theorem that polynomials can be used to approximate a continuous non-linear function arbitrarily well, which explains why models evolved using $G_\mathrm{N}$ can perform well in the modelling task. The prediction and simulation error of the pareto-optimal model from the second experiment with 11 model parameters is plotted in Fig. \ref{fig:CDModelResponse}. The coupled drives case demonstrates that even when prior information is not used to tune the grammar to a specific application, the NARMAX grammar can achieve arbitrarily good approximations of the underlying continuous non-linearity (see \cite{billings2013nonlinear}).

In Tab. \ref{tab:CD}, the results achieved by the proposed method are compared with the results reported in \cite{ayala2014cascaded}, where the authors used an approach based on Differential Evolution (DE) and Neural Networks (NN) to model the coupled drives system. It can be observed that the proposed method achieves comparable results in terms of prediction error, and marginally better results in terms of simulation error.

\begin{table}[]
\centering
\begin{tabular}{c|cc}
\hline
Method                                                     & $\mathrm{RMS}_\mathrm{p}$ & $\mathrm{RMS}_\mathrm{s}$ \\ \hline
Ayala \cite{ayala2014cascaded}                                                     & $4.0 \times 10^{-2}$      & $0.18$                    \\
Proposed (with grammar $G$)                                & $4.0 \times 10^{-2}$      & $0.14$                    \\
\multicolumn{1}{l}{Proposed (with grammar $G_\mathrm{N}$)} & $4.4 \times 10^{-2}$      & $0.14$                   
\end{tabular}
\caption{Comparison of performance measures for the coupled drives case study.}
\label{tab:CD}
\vspace*{-0.4cm}
\end{table}

\subsection{Thermal setup}

\begin{figure}
	\centering
	\includegraphics[width = 0.7\linewidth]{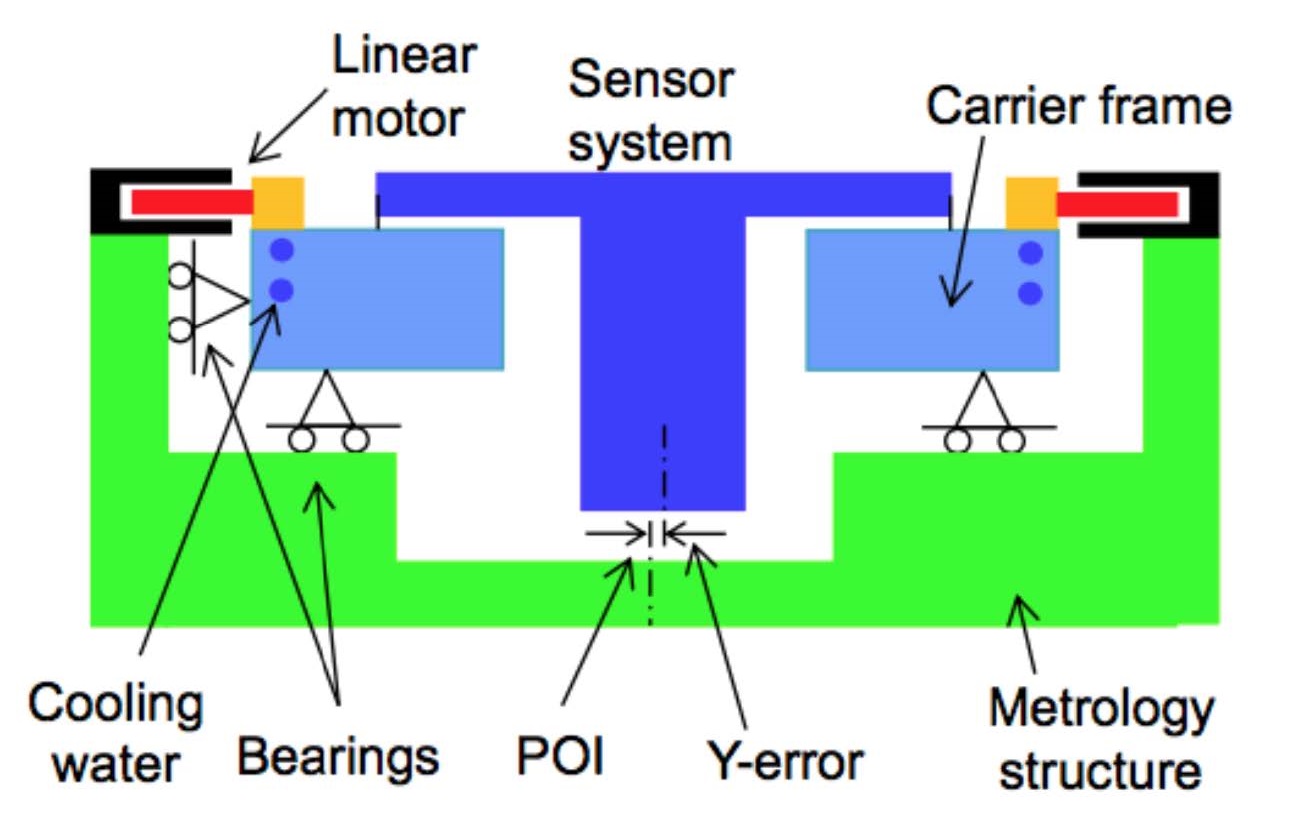}
	\caption{Illustration of the thermal setup.}
	\label{fig:PSA}
\end{figure}

The thermal setup is a positioning system in which a mass is actuated by two motors so as to align a Point-Of-Interest (POI) to a set reference. During actuation, the electric coils of the motors heat up, leading to thermal deformations in the system. For highly accurate positioning, the deformations caused due to the thermal disturbance must be taken into account. A schematic of the system is depicted in Fig. \ref{fig:PSA}. The motor coils cannot be independently actuated, i.e., the same voltage is applied across moth motors coils. The displacement of the POI is measured by a 1-D displacement sensor. For the modelling task, the voltages applied is the excitation signal and the deformations at the POI is the measured output.

The dominant dynamics of the system is due to thermal conduction and convection of heat from the motor coils to the POI and due to thermo-elastic deformations, the former being a slow phenomena. The sampling rate of the system is 1 Hz. The system is excited with a PRBS signal ranging between 0 and 2 V. The deformations are measured in  $\mu$m. Only one dataset of 10810 samples is measured. The same dataset is used for parameter estimation and computing the multi-objective fitness of the models. While the true dynamics of the system are governed by partial differential equations that depend on the geometry of the system, the deformations at a single POI can be well approximated by a simple linear model. The interesting challenge in this case study is the estimation of the delay in the system without any prior knowledge. The NARMAX grammar $G_\mathrm{N}$ is used for this case-study.

\begin{figure}
		\begin{subfigure}[b]{\linewidth}
			\centering
			\includegraphics[scale = 0.96]{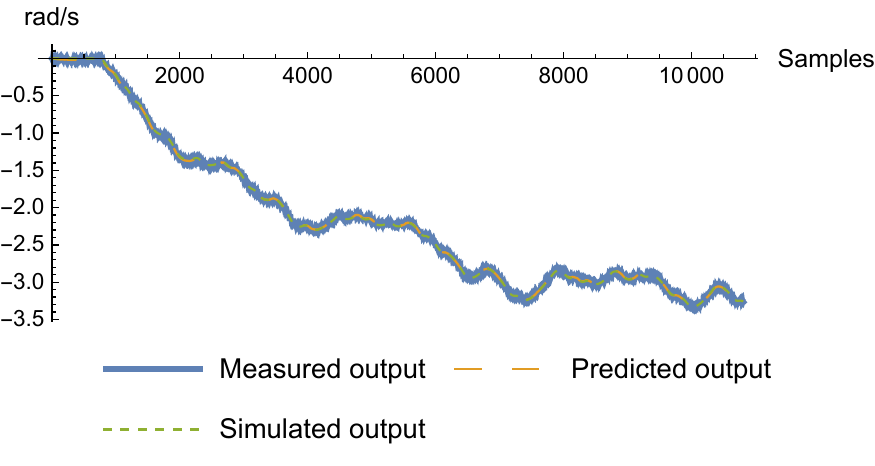}
			\caption{Measured output and the prediction and simulation of the estimated 6 parameter model.}
			\label{fig:PSAModelResponse}
		\end{subfigure}
		\\
		\begin{subfigure}[b]{0.45\linewidth}
		\vspace*{0.2cm}
			\includegraphics[scale = 0.95]{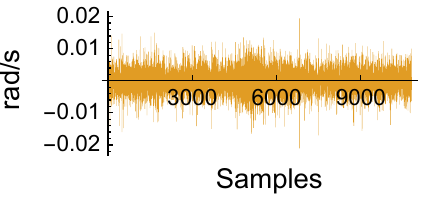}
			\caption{Prediction Error.}
			\label{fig:PSAModelResponsePredErr}
		\end{subfigure}
		~~~~%
		\begin{subfigure}[b]{0.45\linewidth}
			\includegraphics[scale = 0.96]{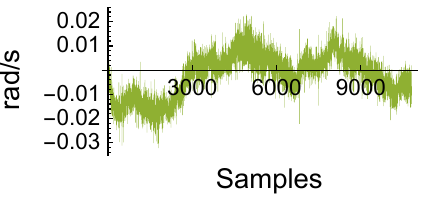}
			\caption{Simulation error.}
			\label{fig:PSAModelResponseSimErr}
		\end{subfigure}
		\vspace*{0.2cm}
		\\
		\begin{subfigure}[b]{0.45\linewidth}
			\includegraphics[scale = 0.95]{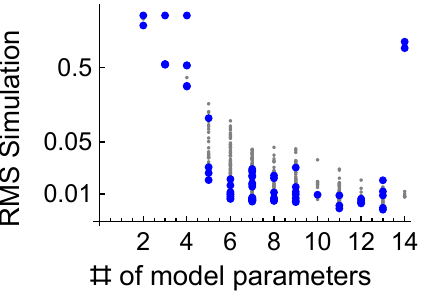}
			\caption{Pareto front - simulation vs complexity.}
			\label{fig:PSASim}
		\end{subfigure}
		~~~~%
		\begin{subfigure}[b]{0.45\linewidth}
			\includegraphics[scale = 0.95]{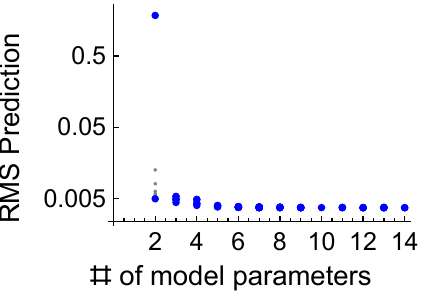}
			\caption{Pareto front - prediction vs complexity.}
			\label{fig:PSAPred}
		\end{subfigure}
		\caption{Numerical results for the thermal case-study.}
		\label{fig:PSAResults}
		\vspace*{-0.4cm}
\end{figure}

The results achieved by the identification algorithm are plotted in Fig. \ref{fig:PSAResults}. The estimated pareto-fronts are plotted in Fig. \ref{fig:PSASim} and \ref{fig:PSAPred}. The models identified by the identification procedure confirms that the dynamics of the system are predominantly linear since most of the models on the optimal pareto-front are linear. Moreover, the identified models were also able to estimate the delay in the system automatically. For e.g., the pareto-optimal model with 6 parameters is given by
\begin{multline}
	y_{k+1} = - 0.00008 u_{k-54} - 0.00014 u_{k-8} + 0.0001 u_{k} - \\
	\quad 0.0377 y_{k-25} +1.0377 y_{k-1} - 0.9628 \xi_{k-1} + \xi_k, \label{eq:PSA}
\end{multline}
which models a delay of 54 samples in the first input term. Note that estimating large delays corresponds to large number of adjunctions of auxiliary tree $\beta_7$ (see Fig. \ref{fig:TAG}) in the model. As a consequence, the identified models also reached the maximum limit of 150 adjunctions during the evolutionary search. Hence, the results obtained could be further improved by increasing the limit of maximum adjunctions for this specific case study. Nonetheless, the results achieved by the proposed identification method are already comparable to state-of-the-art linear identification methods \cite{Ljung1999}. For comparison, the same dataset was used to identify linear models of Output Error (OE) structure using Predicition Error Minimization (PEM) \cite{Ljung1999} method. As an initial step, the delay in the system was estimated (using the \texttt{delayest.m} command in MATLAB). Subsequently, linear models were estimated for a number of model structures, where the delays in input and output terms ranged from 1 to 10. No significant improvement in performance was observed for models with more than 2 poles, corresponding to the model:
\begin{equation}
	y_k = z^{-40}\frac{1.18 \times 10^{-5} - 2.18 \times 10^{-5} z^{-1}}{1 - 1.992 z^{-1} + 0.9918 z^{-2}} + \xi_k.
\end{equation}

A comparison of the best performance measures achieved by the pareto-optimal models identified using the proposed method and the best LTI model (corresponding to 10 poles) achieved using PEM method is presented in Tab. \ref{tab:thermal}. The prediction and simulation output of \eqref{eq:PSA} is plotted in Fig. \ref{fig:PSAModelResponse}. The quality measures are presented in Tab. \ref{tab:overview}.

\begin{table}[]
\begin{tabular}{c|cccc}
\hline
Method   & $\mathrm{RMS}_\mathrm{p}$ & $\mathrm{RMS}_\mathrm{s}$ & $\mathrm{BFR}_\mathrm{p}$ {[}\%{]} & $\mathrm{BFR}_\mathrm{s}$ {[}\%{]} \\ \hline
PEM      & $4.2 \times 10^{-2}$      & $4.2 \times 10^{-2}$      & 99.57                              & 99.57                              \\
Proposed & $3.7 \times 10^{-2}$      & $6.2 \times 10^{-2}$      & 99.63                              & 99.28                             
\end{tabular}
\caption{Comparison of performance measures for the thermal case study.}
\label{tab:thermal}
\vspace*{-0.2cm}
\end{table}


\section{Conclusions}

The modelling framework proposed in \cite{khandelwal2018grammar} was applied to three case studies. The case studies varied in terms of the complexity of dynamics, the amount of data available, the number and accuracy of the sensors and the amount of prior knowledge available. In all three cases, the identification algorithm was used with the almost identical hyper-parameters, except the grammar used. This supports the claim that the use of TAG allows the proposed modelling framework to function across multiple model classes with minimal changes. In the pendulum case study, the identified models successfully captured the dominantly non-linear behaviour of the system. In the benchmark coupled drives case study, the results achieved were comparable to previous results reported in literature. For the thermal setup, the identified results were comparable to that obtained from state-of-the-art linear identification techniques.

\bibliographystyle{IEEEtran}

\bibliography{application_v01}

\end{document}